\documentclass[epjST]{svjour}

\usepackage{graphics}

\begin{document}

\title{Synchronization of a forced self--sustained Duffing oscillator}

\author{Dami\'an H. Zanette\inst{1,2}\fnmsep\thanks{\email{zanette@cab.cnea.gov.ar}}\and Sebasti\'an I. Arroyo\inst{1}}

\institute{Instituto Balseiro and Centro At\'omico Bariloche, 8400
San Carlos de Bariloche, R\'{\i}o Negro, Argentina \and Consejo
Nacional de Investigaciones Cient\'{\i}ficas y T\'ecnicas, Argentina}

\abstract{We study the dynamics of a mechanical oscillator with
linear and cubic forces ---the Duffing oscillator--- subject to a
feedback mechanism that allows the system to sustain autonomous
periodic motion with well--defined amplitude and frequency. First,
we characterize the autonomous motion for both hardening and
softening nonlinearities. Then, we analyze the oscillator's
synchronizability by an external periodic force. We find a regime
where, unexpectedly, the frequency range where synchronized motion
is possible becomes wider as the amplitude of oscillations grows.
This effect of nonlinearities may find application in technological
uses of mechanical Duffing oscillators ---for instance, in  the
design of time--keeping devices at the microscale--- which we
briefly review. }

\maketitle

\section{Introduction}
\label{intro}

Arguably, synchronization is the most basic and most widespread form
of coherent behaviour in interacting dynamical systems
\cite{Strogatz,Pik,Manrubia}. Synchronized dynamics with different
levels of coherence has been observed and characterized in wide
classes of physical, chemical, biological, and social phenomena.
Mathematical, computational, and experimental models have helped to
detect and understand the common elementary mechanisms that drive
synchronization in many  of those systems.  Abstract models of
coupled oscillators have become a very fruitful tool for the
analytical study of coherent evolution in Nature
\cite{Kuramoto,Winfree,HC1,HC2}.

\begin{figure}
\resizebox{.35\columnwidth}{!}{\includegraphics{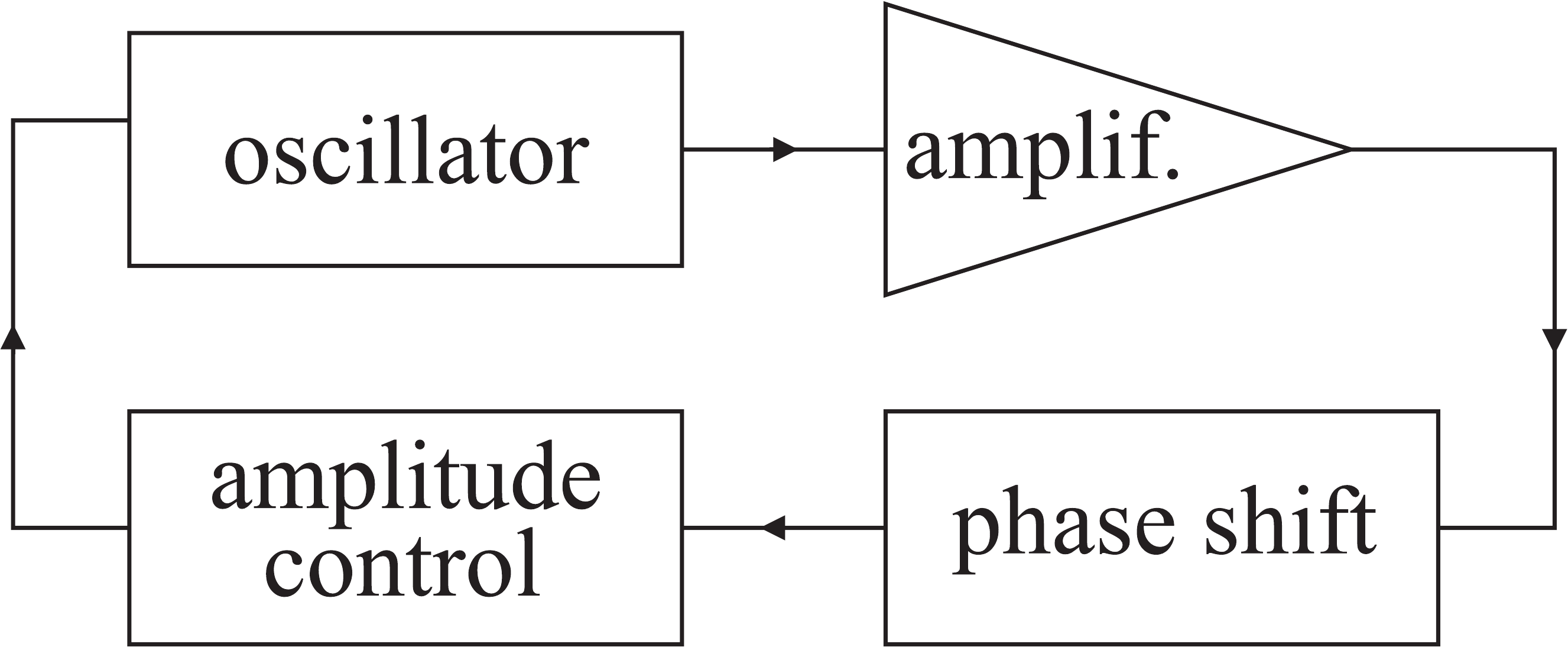}}
\caption{Feedback circuit with signal conditioning for the
self--sustained oscillator. Adapted from Ref.~10.} \label{fig1}
\end{figure}

In the realm of technological applications, electronic elements able
to synchronize the functioning of many components (e.g., clocks) are
present in essentially all devices ---from cell phones and microwave
ovens, to satellites and large power plants.  The need for
miniaturization of electronic circuits has led to considering
replacement of traditional quartz--crystal clocks ---which are
difficult to build and encapsulate at very small scales--- by
micromechanical oscillators \cite{CH,Ek}.  These are minute  silica
elements, that can be directly integrated into circuits during
printing, and  actuated by means of low--power electric fields.  To
generate a sustained periodic signal with autonomously--defined
frequency, they are inserted in a feedback  electronic loop
(Fig.~\ref{fig1}).  In this kind of circuit, the electric signal
read from the  oscillator is amplified and conditioned by, first,
introducing a fixed phase shift and, second,  adjusting its
amplitude to a prescribed value. The conditioned signal is then
reinjected as an external force acting on the oscillator, which thus
responds to its own signal as an ordinary mechanical resonating
system, developing periodic motion with well--defined amplitude and
frequency   \cite{Yurke}.  The only external input on the
self--sustained oscillator is the power needed to condition the
signal; otherwise it acts as an autonomous dynamical system.

In this paper we study the dynamics of a self--sustained mechanical
oscillator driven by elastic and cubic nonlinear forces,
i.e.~governed by the Duffing equation \cite{Nayfeh}. It has since
long been known that the Duffing equation describes the vibrations
of a solid elastic beam clamped at its two ends
\cite{Narashima,Tufillaro}. More recently, it has been
experimentally demonstrated that the motion of a microoscillator
consisting of a clamped--clamped silica beam is well--described by
the same equation, at least, for small to moderately large
oscillation amplitudes \cite{NatComm}. Our main results are obtained
within simplified but realistic approximations, yielding compact and
useful analytical expressions. In Sect.~\ref{sec2}, we find the
frequency and the amplitude of autonomous oscillations in the case
where the phase shift introduced by  signal conditioning maximizes
the oscillator's response to self--sustaining. In particular, we
show that, for hardening and softening nonlinearities, the
oscillation frequency respectively increases and decreases as the
conditioned amplitude grows. In the latter situation, oscillations
are not possible above a certain critical value of that amplitude.
In Sect.~\ref{sec3}, we move to analyze synchronization of the
self--sustained Duffing oscillator under the action of an external
harmonic force.  There, we obtain our most interesting result: under
appropriate conditions, synchronization can be enhanced by making
the amplitude of oscillations larger ---a counterintuitive effect of
nonlinearity. The stability of synchronized motion is assessed
numerically. Finally, we draw our conclusions in the last section.

\section{The self--sustained Duffing oscillator}
\label{sec2}

The motion of a clamped--clamped micromechanical oscillator in its
main oscillation mode is well--described by the Newton equation for
a coordinate $x(t)$  quantifying the oscillator's displacement from
equilibrium \cite{NatComm}:
\begin{equation} \label{osc1}
m \ddot x + \gamma \dot x + k x + k_3 x^3 = F_0 \cos (\phi+\phi_0) +
F_s \cos \Omega_s t.
\end{equation}
Here, $m$, $\gamma$, $k$, and $k_3$ are, respectively,  the
effective mass, damping coefficient, elastic constant and
cubic--force coefficient associated to the dynamics of $x$. Positive
and negative values of $k_3$ correspond, respectively, to hardening
and softening cubic forces. The first term in the right--hand side
represents the  self--sustaining force.   Its amplitude $F_0$ is
fixed by conditioning of the oscillator's signal, as explained in
the Introduction. The angle $\phi$ is the phase associated to the
coordinate during oscillatory motion, $x = A \cos \phi$, and
$\phi_0$ is the phase shift introduced by signal conditioning. The
oscillator's response to the self--sustaining force is maximal for
$\phi_0 = \pi/2$ \cite{NatComm,SADZ}, when the force is in--phase
with the coordinate's velocity $\dot x$. Additionally, we have
included an external harmonic force of amplitude $F_s$ and frequency
$\phi_s$, whose capability of entraining the oscillator into
synchronized motion is studied in Sect.~\ref{sec3}.

Redefining time as a dimensionless variable, $t \sqrt{k/m}  \to t$,
 Eq.~(\ref{osc1}) can be rewritten as
\begin{equation} \label{osc2}
\ddot x+Q^{-1} \dot x + x + \beta x^3 = f_0 \cos (\phi+\phi_0) + f_s
\cos \Omega_s t,
\end{equation}
with $Q=\sqrt{km}/\gamma$, $\beta=k_3/k$, $f_0=F_0/k$, and
$f_s=F_s/k$. Frequencies are now measured in units of $\omega_0 =
\sqrt{k/m}$, the natural frequency of the corresponding autonomous
undamped linear oscillator. Meanwhile,  $f_0$ and $f_s$ have the
same units as the coordinate $x$, and $\beta$  has units of
$x^{-2}$.  Note that the dimensionless quantity $Q$ is the
oscillator's quality factor, which measures the ratio between the
decay time due to damping and the oscillation period. Its inverse
$Q^{-1}$ gives the ratio between the width of the resonance peak and
the resonance frequency.

We first consider Eq.~(\ref{osc2})  for the \textit{unforced}
self--sustained Duffing oscillator ($f_s=0$). In this situation, we
expect that ---due to the action of the self--sustaining
mechanism--- the system asymptotically  attains periodic
oscillations  whose amplitude and frequency are determined by its
own dynamics. Also, for convenience in the analytical treatment, we
fix the phase shift at the value of maximal response:
$\phi_0=\pi/2$.

An approximate harmonic solution to Eq.~(\ref{osc2}) can be found by
applying the standard procedure of neglecting higher--harmonic terms
in the cubic force \cite{Landau} which, in our case, amount to
approximating $\cos^3 \phi \approx \frac{3}{4} \cos \phi$. In the
absence of external forcing,  we propose $x(t) = A_0 \cos \phi
\equiv A_0 \cos \Omega_0 t$, and separate terms proportional to
$\cos  \Omega_0 t$ and $\sin  \Omega_0 t$ to get  algebraic
equations for the oscillation frequency and amplitude.  As shown in
the next section, it is convenient to combine these equations into a
single equation in the complex domain, which reads
\begin{equation} \label{compl0}
(1-\Omega_0^2) A_0 + \frac{3}{4} \beta A_0^3 +i\left( f_0-
\frac{\Omega_0 A_0}{Q} \right) =  0.
\end{equation}
Its solutions are
\begin{equation} \label{hsol}
\Omega_0 = \left( \frac{1+\sqrt{1+3 Q^2 \beta f_0^2 }}{2}
\right)^{1/2}, \ \ \ \ \ A_0 = \frac{Q f_0 }{\Omega_0} .
\end{equation}
Both $\Omega_0$ and the product $|  \beta |^{1/2} A_0$ depend on the
oscillator parameters thorough the combination $\eta = Q|  \beta
|^{1/2}   f_0$ only. Note that $\eta \sim 1$ when the oscillation
amplitude is such that the nonlinear force becomes comparable to the
elastic force, $|\beta| A_0^3 \sim A_0$. It is for those values of
$\eta$ that the frequency begins to  appreciably differ from that of
the linear oscillator ($\omega_0 \equiv 1$). Note also that, for
$\beta <0$ and $1+3 Q^2 \beta f_0^2>0$, an extra solution for the
frequency exists: $\Omega_0 = \left[\frac{1}{2} \left( 1-\sqrt{1+3
Q^2 \beta f_0^2 }\right) \right]^{1/2}$. This solution, however, is
the analytical continuation for $\beta\neq 0$ of a solution with
$\Omega_0 =0$ and $A_0 \to \infty$, and is therefore not expected to
correspond to stable motion.

\begin{figure}
\resizebox{.5\columnwidth}{!}{\includegraphics{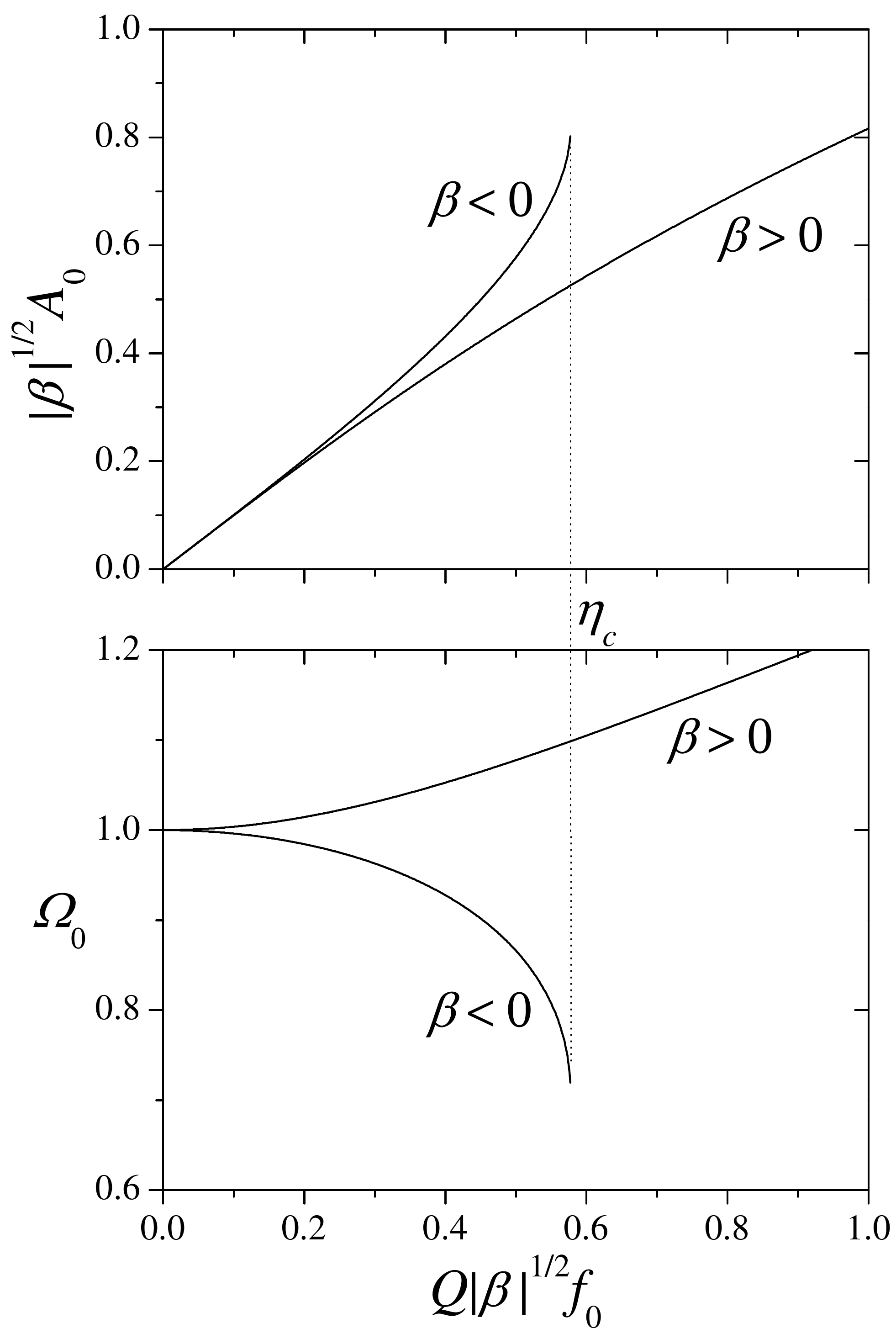}}
\caption{The rescaled oscillation amplitude $|\beta |^{1/2} A_0$ and
the frequency $\Omega_0$ as functions of $\eta = Q|  \beta |^{1/2}
f_0$, for both positive and negative $\beta$. The critical value of
$\eta$, above which the harmonic solution does not exist for
$\beta<0$, is $\eta_c= 3^{-1/2}$.} \label{fig2}
\end{figure}

Figure \ref{fig2} shows the rescaled amplitude $|\beta |^{1/2} A_0$
and the frequency $\Omega_0$ as functions of $\eta$. As could be
expected \cite{Landau}, both for positive and negative $\beta$, the
oscillation amplitude grows when the self--sustaining force and/or
the quality factor increase.   For $\beta >0$, however, the
amplitude growth is sublinear, while it is faster than linear for
$\beta <0$. Respectively, the oscillation frequency increases and
decreases as the amplitude becomes larger.

For $\beta<0$, moreover, the approximate harmonic solution exists
for $\eta \le \eta_c = 3^{-1/2} \approx 0.577$ only. Above this
critical value, the frequency becomes complex, and the solution for
$x(t)$ is no more bounded. The effect of nonlinear softening is too
strong for the system to sustain oscillations of finite amplitude,
and the coordinate grows exponentially.

The stability of the harmonic solution given by Eqs.~(\ref{hsol})
can be assessed along the same lines as for the ordinary forced
Duffing oscillator \cite{Nayfeh}, namely, assuming that damping,
nonlinearities and the external force are perturbations on the
harmonic motion of the free linear oscillator.  The perturbative
calculation, which will be presented elsewhere \cite{fase}, shows
that the harmonic solution for the self--sustained Duffing
oscillator is \textit{globally} stable  ---i.e., it asymptotically
attracts any initial condition---  whenever it exists. Even more,
this result holds whatever the value of the phase shift $\phi_0$ in
the self--sustaining  force.

\section{Synchronized response to harmonic forcing}
\label{sec3}

For the forced self--sustained Duffing oscillator, described by
Eq.~(\ref{osc2}) with $f_s \ne 0$, we seek synchronized solutions
where the coordinate $x(t)$ oscillates with the same frequency as
the external force. Namely, we propose $x(t)=A \cos \phi \equiv A
\cos (\Omega_s t-\phi_s)$, where $\phi_s$ is the (retarded) phase
shift of the coordinate with respect to the force. Replacement in
Eq.~(\ref{osc2}) ---always fixing $\phi_0=\pi/2$, and within the
harmonic approximation for the cubic term--- yields algebraic
equations for the amplitude $A$ and the phase shift $\phi_s$.
Combining them into a  single equation in the complex domain, we get
\begin{equation} \label{compl}
(1-\Omega_s^2) A + \frac{3}{4} \beta A^3 +i\left( f_0-
\frac{\Omega_s A}{Q} \right) =  f_s \exp(-i \phi_s),
\end{equation}
cf.~Eq.~(\ref{compl0}). Under rather general conditions, solutions
to Eq.~(\ref{compl}) exist if the synchronization frequency
$\Omega_s$ lies inside a finite interval  which, as we show below,
also contains the frequency $\Omega_0$ of the unforced oscillator.
This interval is the \textit{synchronization range}
\cite{Pik,Manrubia}.

\begin{figure}
\resizebox{.5\columnwidth}{!}{\includegraphics{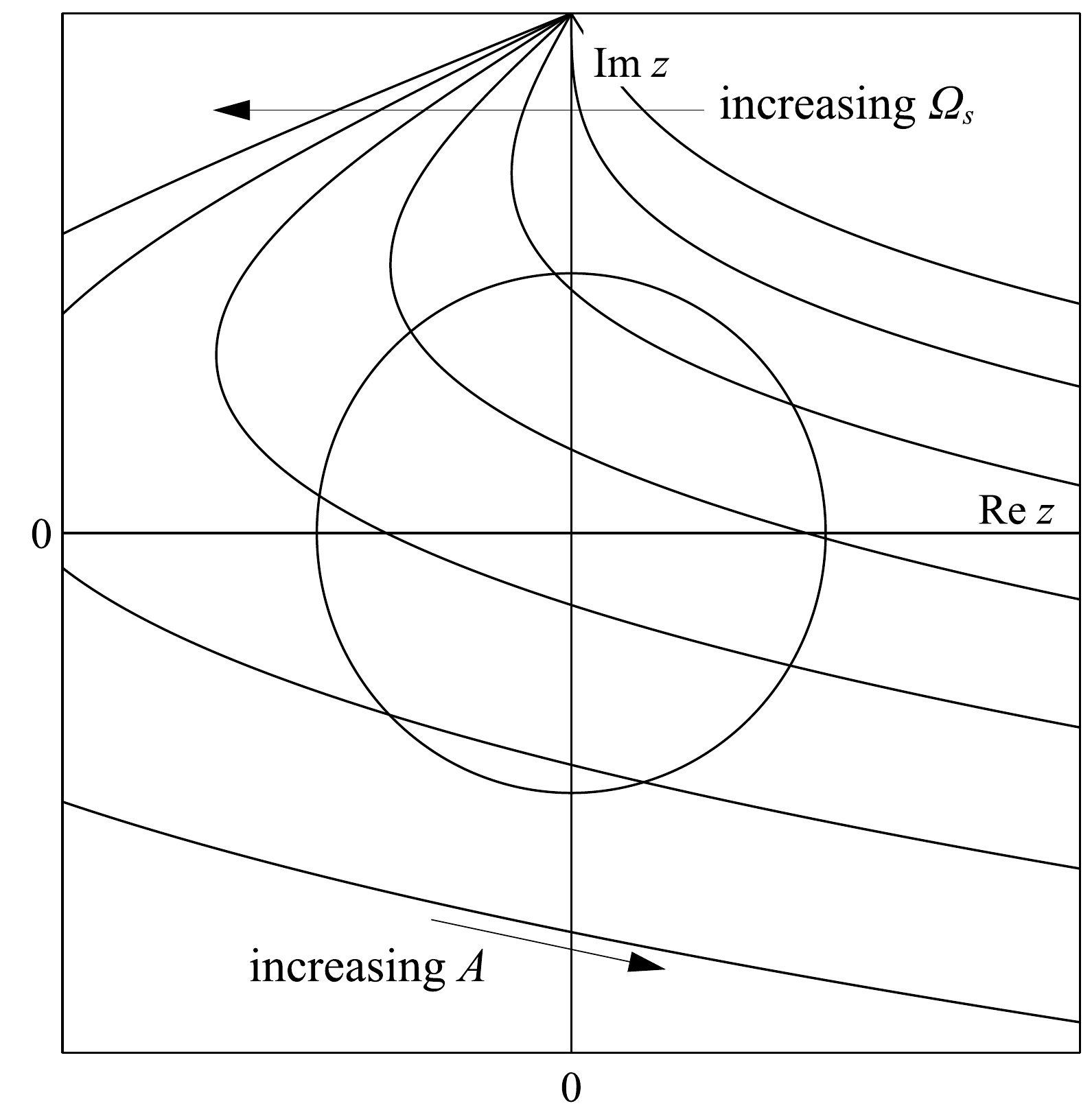}}
\caption{Representation of Eq.~(\ref{compl}) on the complex plane
$z$. The circle centered at the origin represents the right--hand
side of the equation as parametrized by $\phi_s$. Its radius is
$f_s$. The curves represent the left--hand side as parametrized by
$A$, for fixed $\beta$ and $Q$, and several values of $\Omega_s$.
The point on the imaginary axis at which all these curves intersect
each other ($A=0$), is $if_0$.  Along each one of the curves, $A$
increases as indicated by the arrow.} \label{fig3}
\end{figure}

Figure \ref{fig3} illustrates the situation on the complex plane.
The circle centered at the origin, whose radius is $f_s$, represents
the right--hand side of Eq.~(\ref{compl}) as parametrized by the
phase shift $\phi_s$. The curves represent the left--hand side as
parametrized by the amplitude $A$ (with $A\ge 0$), for fixed values
of $\beta$ and $Q$, and several values of the frequency $\Omega_s$.
All these curves pass through the complex number $if_0$, on the
imaginary axis, for $A=0$. Their curvature  is controlled by the
nonlinear coefficient $\beta$, while $\Omega_s$ determines the slope
at $if_0$.  As $\Omega_s$ grows, the curve representing the
left--hand side of Eq.~(\ref{compl}) ``rotates'' around $if_0$ and,
within a finite interval $\Omega_{\min} < \Omega_s < \Omega_{\max}$,
it intersects the circle at two points. Their polar coordinates give
two solutions for $A$ and $\phi_s$. For $\Omega_s = \Omega_{\min}$
or $\Omega_{\max}$ the curve is tangent to the circle, and the two
solutions collapse into a single point.

Note moreover that, by virtue of Eq.~(\ref{compl0}), the curve
representing the left--hand side of Eq.~(\ref{compl}) passes through
the origin of the complex plane for $\Omega_s = \Omega_0$ and
$A=A_0$, given by Eqs.~(\ref{hsol}). Since, in this situation, the
curve necessarily intersects the circle, we conclude that $\Omega_0$
lies within the synchronization range $(\Omega_{\min},
\Omega_{\max})$.

This latter remark suggests that a way to treat Eq.~(\ref{compl})
analytically is to assume that $\Omega_s$ and $A$ respectively
differ from $\Omega_0$ and $A_0$ by perturbatively small quantities.
The limit is achieved for $f_s \to 0$, so that we take as a
perturbative parameter the ratio $p=f_s/f_0$. In the graphical
representation of Fig.~\ref{fig3}, this amounts to taking the
circle's radius much smaller than the distance to the point of
intersection of all the curves on the imaginary axis. Thus, in the
vicinity of the circle, the curves can be conveniently approximated
by straight segments. To implement the approximation, we write
\begin{equation} \label{pert}
\Omega_s = \Omega_0 + p \delta \Omega, \ \ \ \ \
A=A_0 + p \delta A.
\end{equation}
 Expanding Eq.~(\ref{compl}) to the first order in $p$ yields
 \begin{equation} \label{compl1}
 z_0 \delta A -z_1 \delta \Omega =
 \exp(-i \phi_s) ,
 \end{equation}
 with
\begin{equation}
 z_0 = \frac{3Q\beta A_0}{2\Omega_0}-\frac{i}{A_0}, \ \ \ \ \
 z_1= 2Q+ \frac{i}{\Omega_0}
\end{equation}
Unknowns in  Eq.~(\ref{compl1}) are $\delta A$ and $\phi_s$, while
$\delta \Omega$ is given in terms of the parameters of our problem
through the first of Eqs.~(\ref{pert}).  A representation of
Eq.~(\ref{compl1}) on the complex plane, similar to that of
Fig.~\ref{fig3}, makes it immediately possible to show that its
solutions exist when $\delta \Omega$ lies in the interval $(-\delta
\Omega_c, \delta \Omega_c)$, with
\begin{equation}
\delta \Omega_c = \frac{1}{| z_1| |\sin (\zeta_1 -\zeta_0) | }.
\end{equation}
Here, $\zeta_k$ ($k=0,1$) is the argument of the complex number $z_k
\equiv |z_k| \exp (i \zeta_k)$.

For the sake of concreteness, let us analyze this result in the
realistic situation where the oscillator's quality factor is large,
$Q \gg 1$ \cite{NatComm}.
Since ---even when the effects of the nonlinear force
become sizable--- the frequency $\Omega_0$ of the unforced
self--sustained oscillator is expected to remain of order unity, we
can neglect the imaginary part of the complex number $z_1$ by
comparison to its real part, so that  $|z_1|=2Q$ and $\zeta_1=0$.
In this situation, we find
\begin{equation} \label{dW}
\delta \Omega_c = \frac{1}{2Q |\sin \zeta_0|}= \frac{1}{2Q} \left[
\left(   \frac{3Q\beta A_0^2}{2\Omega_0} \right)^2 + 1 \right]^{1/2} .
\end{equation}
Note that, interestingly, this result is independent of the sign of
$\beta$: the domain of existence of synchronized solutions does not
depend on whether nonlinearities are hardening of softening.

\subsection{Analysis of the synchronization range} \label{Asr}

The half--width of the synchronization range, given by the product
$p \delta \Omega_c$  [cf.~the first of Eqs.~(\ref{pert})], depends
on the parameters through the combinations $|\beta|^{1/2} f_0$,
$|\beta|^{1/2} f_s$, and $Q$ only.  The dependence on the two latter
is simple:  $p \delta \Omega_c$ is proportional to $|\beta|^{1/2}
f_s$, and grows monotonically as $Q$ increases with all the other
parameters fixed. As expected, the synchronization range widens when
the external force and/or the oscillator's quality factor become
larger.

\begin{figure}
\resizebox{0.8\columnwidth}{!}{\includegraphics{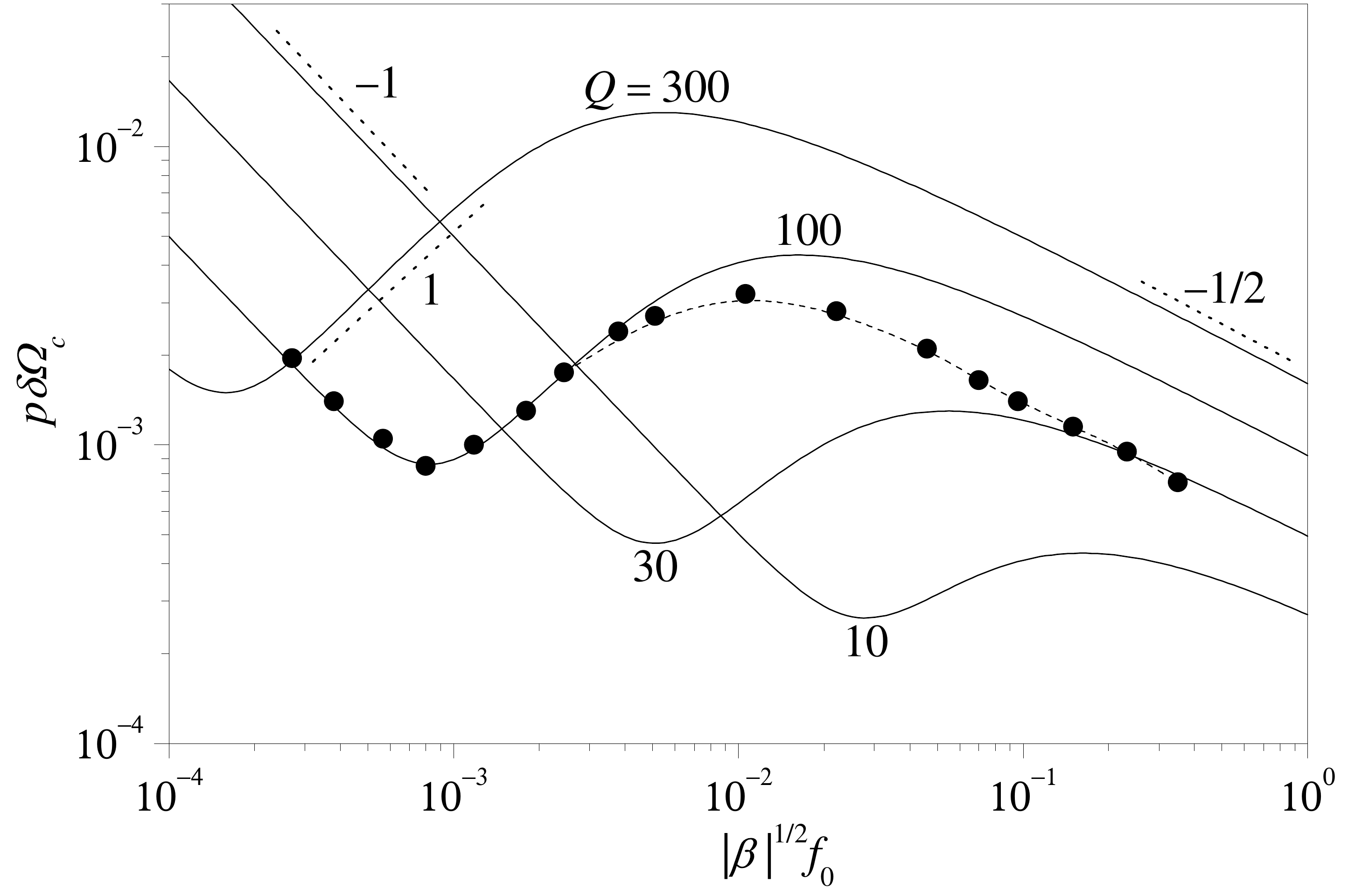}}
\caption{The half--width of the synchronization range, $p \delta
\Omega_c$, as a function of the rescaled self--sustaining force
$|\beta|^{1/2} f_0$, for $|\beta|^{1/2} f_s=10^{-4}$ and several
values of the quality factor $Q$. Dotted labeled straight segments
show the slopes in the three regimes. Full dots stand for numerical
measurements of the synchronization range for $Q=100$. The dotted
curve joining the dots is a spline interpolation plotted as a guide
to the eye.} \label{fig4}
\end{figure}

On the other hand, the dependence of the synchronization range on
the self--sustaining force is less trivial.  When $f_0$ grows, the
amplitude $A_0$ of the self--sustained oscillations increases and,
consequently, we expect that the oscillator becomes more difficult
to entrain by an external harmonic force of  a given amplitude. This
is in fact the standard behaviour of a large class of forced
oscillating systems, including linear oscillators
\cite{Pik,Manrubia,Nayfeh}. For the self--sustained Duffing
oscillator, instead, $p\delta \Omega_c$ exhibits nonmonotonic
behaviour as a function of $|\beta|^{1/2} f_0$. This is shown in
Fig.~\ref{fig4} for several values of $Q$ and $|\beta|^{1/2} f_s =
10^{-4}$.

The dependence of $p\delta \Omega_c$  on $|\beta|^{1/2} f_0$
displays three distinct regimes. For sufficiently small
self--sustaining force, the oscillator operates in the linear
domain. The oscillation amplitude $A_0$ is also small and the
frequency is close to the natural value, $\Omega_0 \approx 1$, such
that $Q|\beta| A_0^2 \ll \Omega_0$.  This inequality can be
rewritten as $|\beta| A_0^3 \ll Q^{-1}\Omega_0 A_0$ which, in the
equation of motion (\ref{osc2}), amounts to having the nonlinear
force $\beta x^3$ much smaller  than the damping force $Q^{-1}\dot
x$. In this limit, the half--width of the synchronization range
reduces to $p\delta \Omega_c \approx f_s/2Qf_0$, and thus decreases
with the amplitude of the self--sustaining force as $f_0^{-1}$.

The second regime is entered when the nonlinear force overcomes
damping, but is still  much smaller than the elastic force: $|\beta|
A_0^3 \ll A_0$. In this situation, the oscillation frequency remains
close to unity, but $\delta \Omega_c$ is now dominated by the first
term inside the square bracket of Eq.~(\ref{dW}). The half--width of
the synchronization range is $p\delta \Omega_c \approx \frac{3}{4}
Q^2 |\beta | f_s f_0$. Counterintuitively, the synchronization range
widens as $f_0$ ---and, consequently, the oscillation amplitude---
grow.

When, finally, the nonlinear force  dominates over both damping and
the elastic force, the variation of $\Omega_0$ with the amplitude
cannot be disregarded anymore. We find that, in this limit of large
self--sustaining force, $p\delta \Omega_c \approx \left( \frac{3}{4}
Q^2 |\beta | \right)^{1/4} f_s f_0^{-1/2}$. Again, as in the first
regime, the half--width of the synchronization range decreases as
$f_0$ grows, now as $f_0^{-1/2}$. The dashed segments in the
log--log plot of Fig.~\ref{fig4}  show the slopes corresponding to
the dependence of $p\delta  \Omega_c $ on $|\beta|^{1/2} f_0$ in the
three regimes.

We remark that the boundaries of the three regimes are completely
determined by the comparison between the rescaled self--sustaining
force $|\beta |^{1/2} f_0$ and a suitable power of the quality
factor $Q$. Namely, the first transition ---from the linear domain
to the intermediate regime--- occurs for $|\beta |^{1/2} f_0 \approx
Q^{-3/2}$, while the second transition takes place for  $|\beta
|^{1/2} f_0 \approx Q^{-1}$. Consequently, as clearly seen in
Fig.~\ref{fig4}, the three regimes become better separated from each
other as $Q$ grows. Note also that, irrespectively of the value of
any other parameter, the oscillator is always in the linear regime
for $\beta=0$.

Taking into account that the above results have been obtained in the
frame of several approximations, it is worthwhile to check their
validity by an independent means. With this aim, we have solved
Eq.~(\ref{osc2}) numerically, to determine within which parameter
ranges are the synchronized solutions actually observed. The
numerical method used to deal with this kind of equation has been
discussed elsewhere \cite{SADZ}. The dots in Fig.~\ref{fig4} stand
for the numerical results for the half--width of the synchronization
range for $Q=100$. We find very good agreement with the analytical
prediction  in the linear regime and in most of the intermediate
regime, while a noticeable departure is apparent in the upper part
of the intermediate regime and for large self-sustaining forces.
This discrepancy may be attributed to at least two sources. First,
the main approximation involved in the analytical results ---namely,
the replacement of the cubic term by a single harmonic function---
is in fact expected to become increasingly inaccurate as the
oscillation amplitude grows. Second, it must be taken into account
that numerical and analytical calculations yield, respectively, the
ranges of \textit{stability} and \textit{existence} of synchronized
motion,  which are not necessarily coincident \cite{SADZ}.   While
we  expect that one of the synchronized solutions is stable within
its whole existence range,  we  cannot discard that its
observability  is jeopardized by the proximity of an unstable
solution (see next section), which would lead the system to converge
to unsynchronized motion from most initial conditions.  Let us
emphasize that, in any case,  our numerical results confirm the
nontrivial behaviour of the synchronization range and, in
particular, the presence of an intermediate regime of
synchronization enhancement, where the synchronization range widens
as the amplitude of oscillations grows.

\subsection{Amplitude and phase shift of synchronized motion}

To complete the characterization of motion within the
synchronization range, we now turn the attention to the amplitude
and phase shift of synchronized oscillations. As discussed in
connection to  Eq.~(\ref{compl}), two synchronized solutions exist
inside the synchronization range. The fact that, upon variation of
the frequency $\Omega_s$ of the external forcing, the two solutions
appear at $\Omega_{\min}$ and disappear at $\Omega_{\max}$ through
tangent (i.e., saddle--node) bifurcations, indicates that one of
them is stable and the other unstable.

\begin{figure}
\resizebox{0.8\columnwidth}{!}{\includegraphics{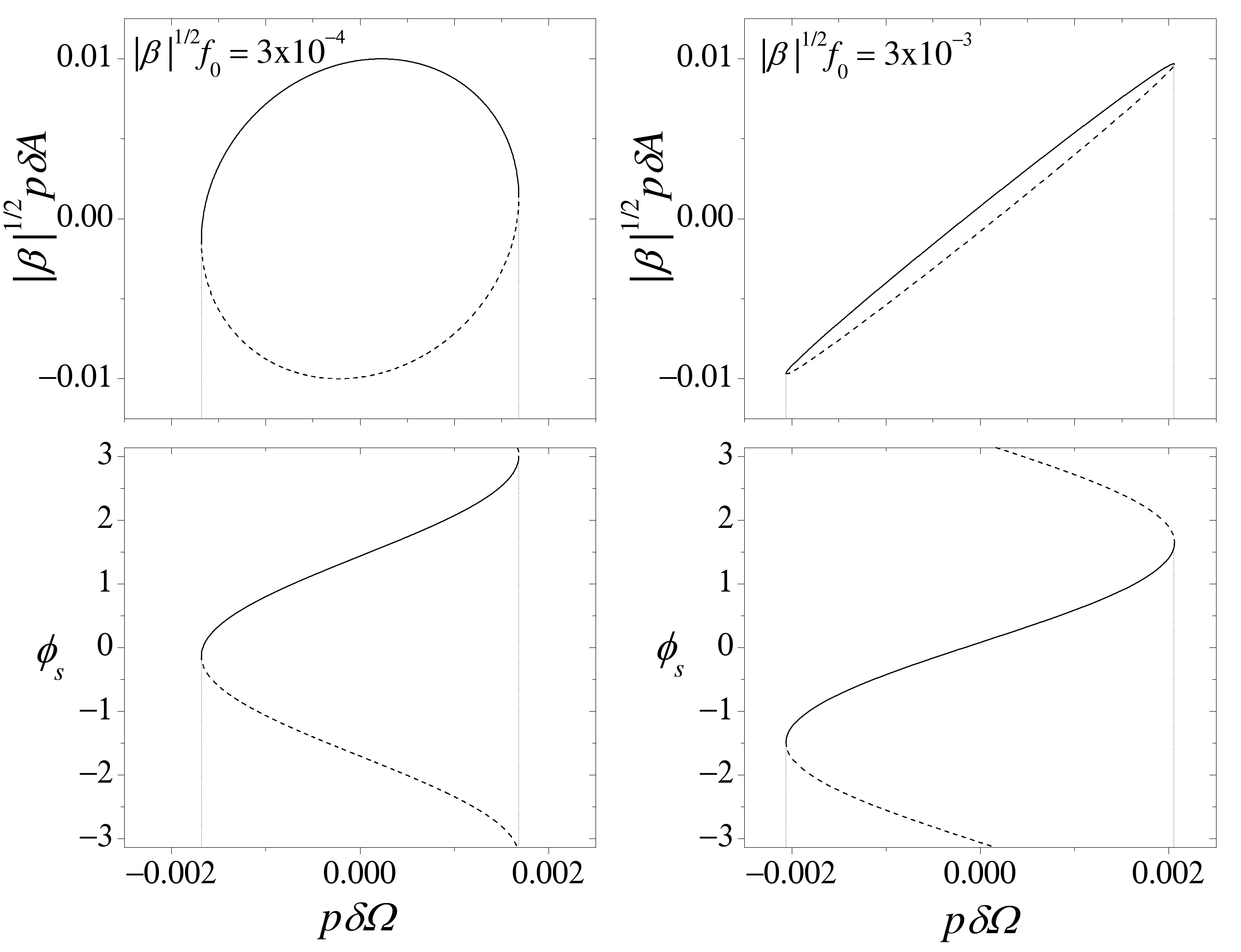}}
\caption{The rescaled amplitude variation $|\beta|^{1/2} p \delta A$
(upper panels) and the phase shift $\phi_s$  (lower panels) of
synchronized motion, as functions of the frequency detuning $p\delta
\Omega$, for $|\beta|^{1/2} f_s=10^{-4}$, $Q=100$, and two
self--sustaining force amplitudes: $|\beta|^{1/2} f_0 = 3\times
10^{-4}$ (left) and $3\times 10^{-3}$ (right). Full and dashed
curves correspond, respectively, to stable and unstable solutions.
The vertical dotted segments indicate the boundaries of the
synchronization range $(-p\delta \Omega_c, p\delta \Omega_c)$.}
\label{fig5}
\end{figure}

Within the approximations considered in the previous section,
squaring Eq.~(\ref{compl1}) establishes a quadratic relation between
$\delta A$ and $\delta \Omega$, which can be  worked out explicitly.
Once $\delta A$ has been obtained, it is reinserted in the equation
to calculate the phase shift $\phi_s$.  Note that, in contrast with
the half--width of the synchronization  range $p \Omega_c$, these
results are \textit{not} independent of the sign of the cubic
coefficient $\beta$. The upper panels of Fig.~\ref{fig5}  show the
rescaled amplitude variation, $|\beta|^{1/2} p \delta A$ as a
function of the detuning  $p\delta \Omega = \Omega_s - \Omega_0$ for
$|\beta|^{1/2} f_s=10^{-4}$, $Q=100$, and two values of
$|\beta|^{1/2} f_0$ in the linear regime and in the intermediate
regime (cf. Fig.~\ref{fig4}). In both cases, $\beta>0$. The lower
panels show the corresponding phase shifts. Full and dashed lines
stand for the stable and unstable solutions, respectively. They
exist within the synchronization range only, whose boundaries are
indicated  by the  vertical dotted segments.

For $|\beta|^{1/2} f_0=3\times 10^{-4}$, the amplitude of the stable
solution first grows with the detuning from a small value at
$-p\delta\Omega_c$, to a maximum just to the right of $p\delta
\Omega=0$. At that point, the oscillator's response to the external
force is maximal.   The phase shift is exactly $\phi_s=\pi/2$, so
that the oscillation velocity and the external force are mutually
in--phase. Moreover,  they are also in--phase with the
self--sustaining signal. As the detuning grows further, the
amplitude decreases until the solutions disappear at $p\Omega_c$.
For $\beta<0$, the behaviour is symmetric  with respect to $p\delta
\Omega=0$. Namely, the maximal amplitude, with $\phi_s=\pi/2$, is
reached to the left of exact tuning.

As $|\beta|^{1/2} f_0$ becomes larger, the ellipse that represents
$|\beta|^{1/2} p \delta A$ as a function of  $p\delta \Omega$
narrows toward a straight segment diagonal to the graph. For
$|\beta|^{1/2} f_0=3\times 10^{-3}$, in the middle of the
intermediate regime, the upper--right panel of Fig.~\ref{fig5} shows
that the maximal amplitude of the stable solution has strongly
shifted to the right, to practically coincide with the upper
boundary of the synchronization range. The phase shift has been
modified accordingly so that we still have $\phi_s=\pi/2$ at the
maximum. This trend persists for larger amplitudes of the
self--sustaining force: in the limit, the amplitudes of the two
solutions lie over the graph's diagonal while, as the detuning grows
within the synchronization range, the phase shift increases from
$-\pi/2$ to $\pi/2$ for the stable solution and decreases from
$3\pi/2$ ($\equiv -\pi/2$) to $\pi/2$ for the unstable solution.

Whereas, as stated above, $\delta A$ and $\phi_s$ can be
analytically calculated as functions of  $\delta \Omega$, the
explicit expressions are too space--consuming to be reported here.
We are however able to provide compact expressions in the linear
limit, $|\beta |^{1/2} f_0 \ll Q^{-3/2}$, and for asymptotically
large self--sustaining force $|\beta |^{1/2} f_0 \gg Q^{-1}$ (see
Sect.~\ref{Asr}). In the former case, we have
\begin{equation}
\delta A = \pm 2Q^2 f_0 \sqrt{1-\left( \frac{\delta \Omega}{2Q}
\right)^2}, \ \ \  \ \ \exp (i\phi_s )= -2Q \delta \Omega \pm
i\sqrt{1-4Q^2\delta \Omega^2},
\end{equation}
where the upper and lower signs correspond, respectively, to the
stable and unstable solutions. Meanwhile, the large--$f_0$ limit
yields
\begin{equation}
\delta A = \frac{2\delta \Omega}{\sqrt{3|\beta|}}  \mbox{  for
$|\delta \Omega | < \delta \Omega_c$}, \ \ \  \ \ \exp (i\phi_s )=
\pm \sqrt{1-\left( \frac{\delta \Omega}{\delta \Omega_c} \right)^2}+
i \frac{\delta \Omega}{\delta \Omega_c},
\end{equation}
with $\delta \Omega_c$ as given in Sect.~\ref{Asr} for the same
limit.

\section{Conclusion}

In this paper, we have characterized the autonomous dynamics and the
synchronizability by a harmonic external force of a self--sustained
Duffing oscillator. The self--sustaining mechanism allows the system
to maintain oscillatory motion with internally defined amplitude and
frequency. As expected, the frequency  increases or decreases as the
amplitude grows, respectively, for hardening and softening
nonlinearities. Synchronization with the external force is possible
when the detuning between the force's and the oscillator's frequency
lies below a certain critical value. Our analysis holds for a
specific phase shift in the self--sustaining mechanism,
corresponding to the expected maximum in the oscillator's resonant
response to the feedback signal. Due to unavoidable experimental
fluctuations in that value, however, it would be worthwhile devoting
future work to relax such condition.

Our most relevant result regards the existence of a regime of
synchronization enhancement where the synchronization range widens
as the amplitude of oscillations becomes larger ---a
counterintuitive effect of nonlinearities, found for intermediate
amplitudes of the self--sustaining force. Interestingly enough, this
regime has already been observed in experiments with micromechanical
oscillators consisting of clamped--clamped silica bars \cite{DD},
although results have not been published yet.   Moreover,  this seems
to be the most natural operation regime of this kind of
micromechanical oscillators with lengths of the order of hundreds of
microns, and quality factors around $10^4$ \cite{NatComm}. For
self--sustaining amplitudes in the linear regime, in fact, the
oscillation amplitudes are too small to provide a signal discernible
from electronic noise. For large amplitudes, on the other hand,
results suggest that the oscillator abandons the parameter region
where it is well described by the Duffing model, and a different
description may prove necessary. A sudden increase
in the size of the synchronization range, that could be related to the
phenomenon described here, has also been recently  reported for
mutually coupled oscillators \cite{Agr}.

It remains to be explored whether the regime of synchronization
enhancement  might be advantageously exploited in applications
where, beyond individually producing a sustained periodic signal,
two or more micromechanical oscillators are expected to synchronize
to each  other. This would be the case in building up a more robust
periodic signal ---able to overcome the effects of electronic and/or
thermal noise--- or in devices where synchronous coherent motion of
many oscillators is required, such as in optical components for
communication systems \cite{CH,Ek}.  The collective dynamics of an
ensemble of coupled self--sustained Duffing oscillators is
\textit{per se} an attractive problem that deserves future
consideration.

\end{document}